\documentclass[iop]{emulateapj}
\usepackage{color, hyperref}

\slugcomment{Accepted for publication in the ApJ}

\shorttitle{Rotation periods in M\,67}
\shortauthors{Barnes, Weingrill, Fritzewski, Strassmeier, Platais}

\begin{document}

\title{Rotation periods for cool stars in the 4\,Gyr-old open cluster M\,67,\\ the solar-stellar connection, and the\\ applicability of gyrochronology to at least solar age}



\author{Sydney A. Barnes\altaffilmark{1}, Joerg Weingrill, Dario Fritzewski, Klaus G. Strassmeier}
\affil{Leibniz Institute for Astrophysics (AIP), Potsdam, Germany}

\email{Email: sbarnes@aip.de}

\and

\author{Imants Platais}
\affil{Department of Astronomy, Johns Hopkins University, Baltimore, MD, USA}

\altaffiltext{1}{Also, Space Science Institute, Boulder, CO, USA}

\begin{abstract}
We report rotation periods for 20 cool (FGK) main sequence member stars of the 
4\,Gyr-old open cluster M\,67 (= NGC\,2682), obtained by analysing data from 
Campaign\,5 of the K2 mission with the {\em Kepler} Space Telescope.
The rotation periods delineate a sequence in the color-period diagram (CPD)
of increasing period with redder color.
This sequence represents a cross-section at the cluster age of the
surface $P = P (t, M)$, suggested in prior work to extend to at least solar age.
The current Sun is located marginally (approx. one sigma) 
above M\,67 in the CPD, as its relative age leads us to expect, 
and lies on the $P = P (t, M)$ surface to within measurement precision.
We therefore conclude that the solar rotation rate is normal, as compared
with cluster stars, a fact which strengthens the solar-stellar connection.
The agreement between the M\,67 rotation period measurements and prior 
predictions further implies that rotation periods, especially when coupled with 
appropriate supporting work such as spectroscopy, can provide reliable ages via 
gyrochronology for other similar FGK dwarfs from the early main sequence to 
solar age and likely till the main sequence turnoff.
The M\,67 rotators have a rotational age of 4.2\,Gyr, with a standard deviation 
of 0.7\,Gyr, implying that similar field stars can be age-dated to precisions 
of $\sim$17\%.
The rotational age of the M\,67 cluster as a whole is therefore $4.2$\,Gyr, 
but with a lower (averaged) uncertainty of 0.2\,Gyr.
\end{abstract}


\section{Introduction}

Our Sun is the archetype for a vast plurality of stars, particularly main
sequence cool stars , i.e. those with surface convection zones \citep{Schatzman62}\footnote{Massive stars with radiative surfaces ($M/M_{\odot} > 1.3$) are known to have different behaviors \citep[e.g. ]{Bethe39, Kraft67}, as are stars that have evolved off the main sequence.}.
This idea is enshrined in a Copernican principle whose origins go back to 
ancient Greece, and in research circles today is called 
``the solar-stellar connection'' 
 \citep[e.g. ][]{Mihalas81,Sofia87,Strassmeier04,Brun15}.

Various pillars of this solar-stellar connection include the consonance between
the Sun on the one hand, and similar stars in the Galaxy on the other hand, 
in terms of e.g. chromospheric activity levels 
\citep{Eberhard13,Hale04, Wilson63},
the lengths of their magnetic cycles \citep{Wilson63,Baliunas95},
photometric variability \citep{Lockwood07,Lockwood13},
and characteristics of their starspots 
\citep{Kron47,Vogt83,Strassmeier09}. 

The chromospheric activity of solar-type stars in particular has been carefully 
studied over the years, and is known from open cluster work to decline 
systematically enough with stellar age to be used as a reliable, if coarse, 
age indicator \citep[e.g.][]{Wilson78,Skumanich72,Noyes84,Soderblom91,Donahue98, Mamajek08}. 
However, the large distances to old clusters coupled with meager chromospheric 
emission from old stars mean that we are forced to work with nearby field stars,
 such as the Mt. Wilson sample \citep[e.g.]{Baliunas96}, whose ages are not 
independently known\footnote{The age of the Sun itself (and the solar system) is known from radioactive dating of meteoritic material to be $4.57$\,Gyr \citep{Patterson56,Allegre95}.}. 
This has made it difficult to assess the activity of {\em solar-age} stars.
To clarify this part of the solar-stellar connection, \citep{Giampapa06} 
studied chromospheric activity in solar-type stars of M\,67 and found both
a mean activity level resembling solar values, and a wide activity range,
the latter suggesting caution, perhaps even pessimism, in using this activity 
routinely as an age indicator.

The solar-stellar connection is also invoked in models of stellar spindown
\citep[e.g.][]{Endal81, Kawaler88, Pinsonneault89} because
these both reasonably assume that the angular momentum loss from the magnetized 
solar wind \citep{Parker58, Weber67} is a good model for other cool 
star winds, and also calibrate the efficiency of angular momentum loss by 
requiring that a solar-mass stellar model has the solar rotation period 
($26.09$\,d at the average sunspot latitude; \citet{Donahue96}) 
at solar age.

In terms of rotation itself, data acquired steadily over the last 50 years 
have also shown, initially that the rotation velocities, and since the 1980s, 
that the less ambiguous rotation periods, $P$, of late-type main sequence stars 
are largely dependent on  stellar age, $t$, and mass, $M$, or other suitable 
mass proxy such as $(\bv)_0$ color 
\citep{Kraft67, Skumanich72, vanLeeuwen82, Radick87, Kawaler89, Barnes03, Barnes07, Meibom11, Meibom15}.
This relationship can be represented by a surface $P = P (t, M)$ in the
corresponding three dimensional ($P, t, M$) space, which can then be inverted 
analytically or numerically to provide the star's age $t = t (P, M)$, 
a procedure known as gyrochronology \citep{Barnes03,Barnes07,Barnes10}. 

A key observational goal has been to define the $P = P (t, M)$ surface 
empirically to the oldest-possible ages by measuring rotation periods for 
cool cluster member stars in a succession of open clusters, each cluster 
sampling a slice of this surface at constant age. Clusters, after all, provide 
homogeneous and coeval stellar samples whose ages -  isochrone or otherwise - 
are hugely more reliable than those of field stars 
\citep[e.g.][]{Sandage62, Demarque64}.
{\em Kepler} observations of key open clusters, coupled with onerous but 
essential membership and multiplicity surveys, have already extended the known 
surface beyond the 625\,Myr age of the Hyades \citep{Radick87, Delorme11}, first to 1\,Gyr \citep[NGC\,6811;][]{Meibom11}, 
and recently to 2.5\,Gyr \citep[NGC\,6819;][]{Meibom15}. 

The Sun apparently lies on the same surface as that defined by the open 
clusters, but because the oldest available cluster to date (NGC\,6819) is still
$\sim$2\,Gyr younger than it, there is room for ambiguity.
K2 observations of the open cluster M\,67, believed to be $\sim$4\,Gyr old 
\citep{Demarque92,VandenBerg04, Bellini10},
make it possible to extend the $P = P (t, M)$ surface out to near-solar age, 
and to assess to what extent the Sun does (or does not) lie on it.
If it does, then the $(P, t, M)$ relationship is likely valid for the entire
main sequence until the turnoff, as the rotation periods of the Mt.\,Wilson 
field star sample seem to suggest 
\citep[see e.g.][]{Baliunas96,Barnes03,Barnes07}.
This issue is at the heart of the solar-stellar connection.
It is addressed here by providing rotation periods for 20 cool (FGK) stars in 
M\,67.

\section{Data analysis}

\begin{figure}
\includegraphics[trim=90 110 80 150,clip=true,width=\columnwidth]{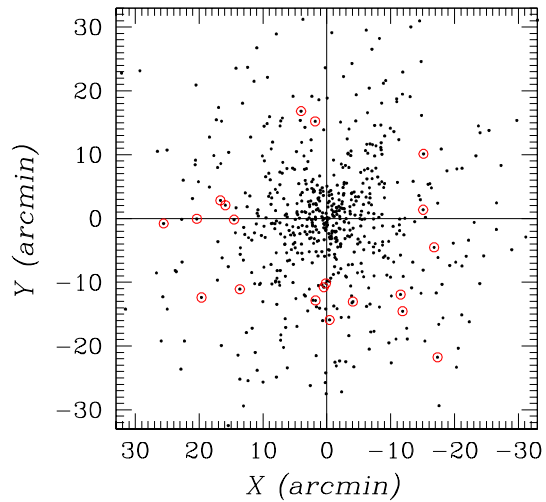}
\caption{\label{figure1}Spatial distribution of known line-of-sight velocity members with $P_{\rm RV}\ge50\%$ (black dots). Our target stars are marked with red circles. $X, Y$ coordinates are from the gnomonic projection of equatorial coordinates with respect to the adopted center of M67 ($\alpha = 8^{\rm h}\,51^{\rm m}\,23.3^{\rm s}, \delta = +11^{\circ}\,49'\,02''$, J2000; GLM15).
}
\end{figure}

We downloaded the K2 (Campaign\,5) time-series data for all available main 
sequence targets from the MAST archive\footnote{\url{https://archive.stsci.edu/k2/}}. 
We ignored the crowded ($\sim 1/4^{\circ}$ square) central region of the cluster 
that is covered by a K2 superstamp, and concentrated on the light curves for 
those individual K2 targets that are located in the annular region beyond this 
limit but within $1/2^{\circ}$ from the center, and that are listed in the 
recent M\,67 membership study of \citet[hereafter GLM15]{Geller15}. 
Fig.\,\ref{figure1} displays the spatial locations of all GLM15 member stars 
(with radial velocity membership probability greater than 50\%) 
in the M\,67 region of the sky. 
We began with all released K2 stars listed in GLM15, regardless of their 
membership status to enable us to remove spacecraft, instrumental, and 
non-astrophysical signatures from the K2 light curves.

Although the presearch data conditioning (PDC) light curves provided by the K2 
C5 data release have removed many spacecraft glitches, global study of 
all available light curves clearly shows residual instrumental and/or 
spacecraft signatures that also need to be removed\footnote{ The data release notes are available at \url{http://keplerscience.arc.nasa.gov/K2/C5drn.shtml}. See also \citet{VanCleve15}.}.
We performed principal component analysis (PCA), with all lightcurves
matched against epochs, as specified by the K2 flag {\sc cadence\_no}.
PCA immediately shows that there are two major families of light curves that 
originate in the M\,67 stars being located on two separate CCDs (`channels'
in K2 jargon). 
These two groups follow each other relatively closely at the $\sim$5\,mmag 
level, but have differing characteristics below this level, and were therefore 
corrected separately. Spacecraft drift is also visible in the PCA.
There is also a small number of `light curves' with almost no signal.
These were ignored in subsequent analysis.

We subtracted the mean flux for each light curve, and also the median for each 
epoch, representing the central tendency of the entire ensemble of light curves.
(This is equivalent to putting all stars on the same photometric system, 
normalizing each star's flux to its mean value, and removing the common trends.)
This procedure decreased the RMS for the individual channels by a factor of
$\sim$5, with the best-exposed light curves for the inactive stars at this 
stage giving an RMS of 0.3\,mmag over the whole light curve. 
(The presence of such `constant stars' informs us that data corrections are
 successful to this level, and that variability above this level for comparably
 bright stars is likely intrinsic.)
We also removed all data points with K2 flag {\sc sap\_quality} greater than 
zero, and followed it by linearly interpolating single missing datapoints 
\citep{Weingrill15}, to produce equidistant time steps, 
enabling the application of a low-pass one-half-day filter, and the application
of the regular fast Fourier transform.
This procedure does not affect the power spectrum, as explained in \citep{Weingrill15}, which describes the overall software framework for our studies, and
which originates in software developed to reduce CoRoT satellite data.
Many light curves and power spectra show features from the ``Argabrightening'' 
event 38\,d into C5. There are also residual effects from a coronal mass
ejection at 55.5\,d from C5 start.
Data from the first and last days of K2 C5 observations were also
found to be unstable enough that they had to be discarded. 
The final dataset nominally consists of a 73\,d time series.

Several high-amplitude variables (binaries etc.) are recognizable 
directly from the PCA analysis performed on all downloaded stars. However,
these are not relevant to our aims, being either non-members or too bright, 
and were ignored.
The period search algorithms  were restricted to the 106 single and binary 
cluster members in the region of the CMD below the turnoff at $V=13.5$.
Pulsators are not found in our region of the CMD (late-F to mid-K), 
and in any case would be easily identifiable from their power spectra.
The light curves of a significant fraction ($\sim$30\%) of the cool single and
binary cluster members display distinctive mmag-level light curve modulations 
that are (and have long been) recognizable from ground-based and space 
observations as originating in star spots. We observe activity, and also 
significant evolution of the spots over the 75\,d baseline of K2 observations.
Rotation periods were derived for these spotted variables using four principal
methods, phase dispersion minimization \citep[PDM;][]{Stellingwerf78},
minimum string length \citep[SL;][]{Dworetsky83}, 
the Bayesian period signal detection method 
\citep[GL;][]{Gregory92, Gregory99}, 
and auto-correlation \citep[AC;][]{Scargle89}. 
None of these make prior assumptions about the shape of the light curve, 
a fact that will be important in the sequel. 
We also used the fast Fourier transform (FFT)\footnote{For the FFT analysis, we used zero padding and a Hamming window to dampen possible aliases and ringing effects.}. 
However, because most of the light curves are not sinusoidal, 
 (70\% have two major spot groups; see Table\,1)
the FFT sometimes preferentially locks onto roughly half the true period,
and was downplayed in this particular context.
The domain for the period searches was set to $0.6$\,d$ < P < 37$\,d, 
with the long-period limit specified by our requirement that at least two 
phases be seen for each variable in the 75\,d K2 observing window.
Our period resolution is nominally 0.05\,d.
We generally sought period agreement between all four of the PDM, SL, GL, and 
AC methods to retain a given star, but all methods are not universally 
successful. The baseline of the K2 observations is long enough for significant 
spot evolution to be observed in most of the stars, and yet not long enough to 
provide multiple phases for these long-period stars. 
The final decision was made manually, guided by the noise properties of the 
individual light curves. 
We demanded that all accepted light curves have a dip in the PDM
 $\Theta$ statistic below $\Theta = 0.55$, that this periodicity be obtained 
  from at least two of the other methods, and that the light curves should 
  demonstrate this variability subjectively upon visual examination.

As a result of the above procedure, 
we settled on 19 single and 1 binary cluster member stars displaying 
periodic spot-induced variability, and these will be our sole concern below\footnote {Although additional stars could be added by relaxing our criteria - 5 more by allowing $0.55 < \Theta < 0.75$, we conservatively decided not to include them in this first paper. These stars also lie on the same sequence in the cluster CPD, and do not alter our results.}. 
The locations of these 20 stars on the sky are marked in Fig.\,1, together
with those of the other member stars in GLM15 with radial velocity membership
probability $P_{\rm RV} \ge 50\%$. The annular distribution of the sample,
as mentioned earlier, is clearly visible.

The light curves for all stars are displayed in the accompanying online 
appendix, 
together with the associated smoothed `power spectra' for the four principal 
period search methods used.
The variability levels for the solar-type stars reported here are similar to 
that of the active Sun, and other stars in our sample have variability
similar to that of the quiet Sun \citep{Lockwood07}.
The final chosen period and its error, calculated from Gaussian fits to the
smoothed PDM spectra, are displayed in each power spectrum panel.  
70\% (14/20) of the light curves show the presence of two large spot groups.
The rotation periods themselves, together with their uncertainties 
and other stellar information relevant to this paper are listed in Table\,1. 

\begin{deluxetable*}{crccccccl}[b!]
\tablewidth{0pt}
\tablecaption{Data table for M67 member stars with rotation periods}
\tablenum{1}
\tablehead{\colhead{EPIC} & \colhead{IDW} & \colhead{$\bv$} &
\colhead{V} & \colhead{P} & \colhead{P$_{\rm err}$} & \colhead{Groups} &
\colhead{Member} & \colhead{Comment} \\
\colhead{} & \colhead{} & \colhead{} & \colhead{(mag)} &
\colhead{(days)} & \colhead{(days)} & \colhead{} & \colhead{} & \colhead{} }
\startdata
211388204 &  8056 & 0.79 & 15.05 & 31.8 & 1.6 & 2  & SM & $15.6\arcsec$, $+$3.01\\
211394185 & 16032 & 1.04 & 15.69 & 30.4 & 1.8 & 2  & SM & ...\\
211395620 & 27038 & 1.02 & 16.35 & 30.7 & 3.6 & 1  & SM & $16.5\arcsec$, $-$2.6; Not in K2\\
211397319 & 18028 & 0.67 & 14.57 & 25.1 & 1.3 & 2  & SM & $13.5\arcsec$, $-$0.43 = 211397127\\
211397512 & 26026 & 1.00 & 16.18 & 34.5 & 2.2 & 2  & SM & $11.6\arcsec$, $+$4.99\\
211398025 & 13047 & 0.78 & 15.25 & 28.8 & 1.6 & 2  & SM & $11.8\arcsec$, $-$1.34 = 211398328\\
211398541 & 19034 & 0.98 & 16.20 & 30.3 & 6.8 & 2  & SM & $08.4\arcsec$, $+$2.52\\
211399458 & 25036 & 1.02 & 16.21 & 30.2 & 1.9 & 2  & SM & ...\\
211399819 & 24022 & 0.79 & 15.21 & 28.4 & 2.0 & 2  & SM & ...\\
211400500 & 13021 & 0.67 & 14.53 & 26.9 & 1.5 & -- & SM & ...\\
211406596 & 21035 & 0.82 & 15.16 & 26.9 & 2.2 & 2  & SM & ...\\
211410757 &  8052 & 0.56 & 13.67 & 18.9 & 0.8 & 1  & SM & ...\\
211411477 & 12030 & 0.70 & 14.66 & 31.2 & 1.9 & 2  & SM & $14.8\arcsec$, $+$0.27 = 211411722\\
211411621 &  9041 & 0.82 & 14.66 & 30.5 & 1.0 & 2  & SM & ...\\
211413212 &  8031 & 0.66 & 14.52 & 24.4 & 2.5 & 1  & SM & ...\\
211413961 & 17033 & 0.79 & 15.04 & 31.4 & 1.5 & 2  & SM & $16.2\arcsec$, $+$3.83\\
211414799 &  4034 & 0.58 & 13.71 & 18.1 & 0.5 & 2  & BM & SB1\\
211423010 &  9037 & 0.77 & 15.05 & 24.9 & 1.9 & 2  & SM & ...\\
211428580 & 12031 & 0.74 & 14.74 & 26.9 & 2.3 & 1  & SM & ...\\
211430274 & 28035 & 0.93 & 15.74 & 31.1 & 2.7 & -- & SM & ...\\
\enddata
\tablecomments{IDW is the Identification number from GLM15.
The Member column indicates whether the star is a GLM15 single cluster member 
(SM) or a binary cluster member (BM). 
Column heading Groups indicates the number of large spot groups on the stellar 
surface. 
Stars with brighter ($-$) or fainter ($+$) neighbors within $20 \arcsec$ are 
noted, along with their magnitude difference.
}
\end{deluxetable*}

Cluster membership probabilities are drawn from GLM15, providing both the 
line-of-sight (LOS) membership probabilities, $P_{\rm RV}$, and the compiled 
proper-motion membership probabilities, $P_{\mu}$. 
All LOS membership probabilities of our K2 targets have $P_{\rm RV} \geq 95\%$. 
A total of 16 K2 stars have $P_{\mu} > 75\%$, while 3 additional stars have 
$75\% > P_{\mu} > 50\%$. 
Star IDW 27038 only has a $P_{\rm RV}$ estimate.
Essentially, all 20 stars can be considered as bona fide cluster members.

We checked both the full-frame K2 images and those from the Digitized Sky 
Survey (DSS1), to examine whether any of our light curves might be contaminated 
with light from neighbors.
These can be seen in cutouts from both DSS1 and K2 images of a 
$30''\times30''$ region centered on each star, and are included in the 
appendix to this paper. Note that each K2 pixel is $3.98''$.
The nearest neighbor, at a separation of $\sim8\arcsec$, is 2.5 magnitudes
fainter than the K2 target EPIC 211398541 (= IDW 19034). 
Brighter neighbors can be found at distances exceeding $13\arcsec$, 
and should not make significant contributions to the target's light curve.
Indeed, three bright neighbors (see Table\,1) are K2 targets in their own right,
and their light curves are very different from those of our periodic stars, 
validating our claim.

\section{Results}

\subsection{The rotators in the color-magnitude diagram}

\begin{figure}
\includegraphics[angle=-90,trim=0cm 3cm 0cm 0cm,clip=true,width=100mm]{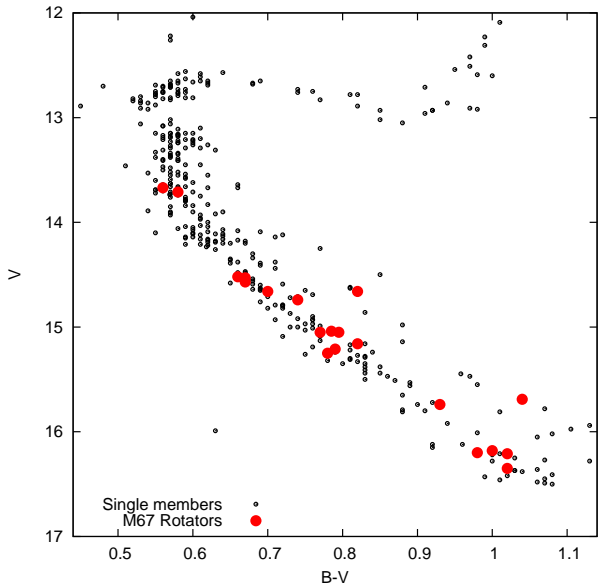}
\caption{\label{figure2}Color-magnitude diagram for M\,67, with the periodic rotators
highlighted (red). 
Only the single cluster members from GLM15 are displayed (black). 
The periodic spotted stars occupy the range 
$0.58 < \bv < 1.04$ i.e. $1.15 < M/M_{\odot} < 0.8$, on the main sequence, 
(The rotator at $\bv = 0.58$ is the only binary in this sample. 
 Two coincident stars at $\bv = 0.79$ are artificially separated in color. 
 The two stars on the photometric equal-mass binary sequence are 
 spectroscopically single cluster members.)
}
\end{figure}

M\,67 has a well-known and rich color-magnitude diagram (CMD), which we 
reproduce in Fig.\,\ref{figure2}, displaying only the single cluster members from 
GLM15. The ($\bv , V$) data plotted here are taken from GLM15, sourced 
mainly but not exclusively from \citet{Montgomery93}. 
\citep[See also][]{Sandquist04, Yadav08}.
The displayed region includes a richly-populated main sequence which is the 
region of interest here, photometric binaries, the turnoff, subgiants, 
and the base of the giant branch.
The 19 single and 1 binary (at $\bv = 0.58$) cluster members for which we have 
derived rotation periods are highlighted in this figure.
The periodic rotators all lie on the cluster's single or photometric binary 
sequence, as expected. They cover the range $0.54 < (\bv)_0 < 1.00$, equivalent
to $1.15 > M/M_{\odot} > 0.8$, encompassing stars with spectral types ranging 
from F8 to K4.
We have used $E(\bv) = 0.04$, as recommended in the exhaustive study of
Taylor (2007).

\subsection{The color-period diagram}

\begin{figure}
\includegraphics[angle=-90,trim=0cm 3cm 0cm 0cm,clip=true,width=100mm]{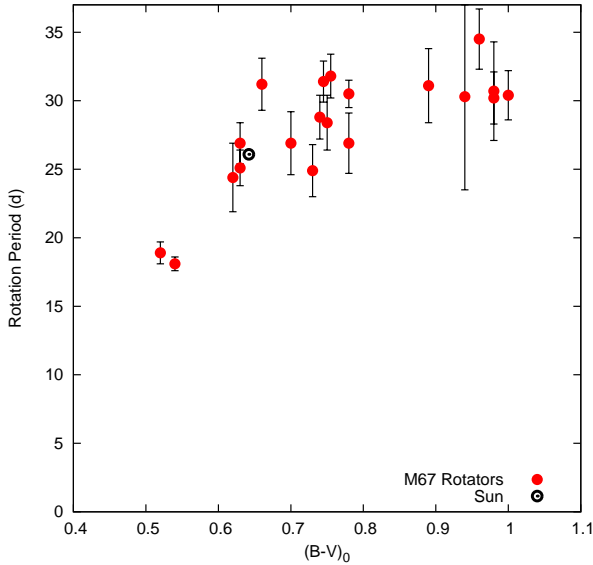}
\caption{\label{figure3}Color-period diagram (CPD) for M\,67, displaying the location of the
20 periodic rotators derived from the K2 data. 
The stars trace a sequence from blue colors and short periods to redder colors 
and long periods, echoing similar shapes seen in younger open clusters at 
much shorter periods. 
The Sun is almost indistinguishable from other M\,67 stars in this diagram.
Two coincident stars at $(\bv)_0 = 0.75$ are artificially separated in color. 
The rotator at $(\bv)_0 = 0.54$ is the only one classified as binary.
}
\end{figure}

The location of the 20 periodic cluster member stars in the corresponding
color-period diagram (CPD) is displayed in Fig.\,3.
The periods are observed to be a function of stellar mass, with the 
bluer/higher-mass stars having shorter periods, and lower-mass stars having 
longer periods.
In fact, although the relationship seen in Fig.\,\ref{figure3} is less tight than e.g. 
the equivalent one in the 2.5\,Gyr-old cluster NGC\,6819 \citep{Meibom15}, 
the stars clearly follow the same trend 
of increasing  period with redder color (decreasing mass)
as the one observed initially in the 
Hyades \citep{Radick87}, and subsequently in a succession of other 
well-studied open clusters, both younger and older, and even field stars, as 
proposed in \citet{Barnes03, Barnes07}. 
In these publications, the shape of the mass dependence was proposed to be 
age-independent and represented by a function, $f = f (B-V)$, only of color. 
Although this was an adequate approximation a decade ago, it is not believed 
to be a completely satisfactory description at this time. 
See B10 for a more recent viewpoint. 

These M\,67 CPD data represent the cross-section of the $P = P (t, M)$ surface 
at the cluster age, $t \approx 4$\,Gyr.
The present-day Sun is very close to the mean curve defined by the M\,67 data 
points, and lies marginaly above the mean- and median isochrones (described 
below), implying that the Sun is likely slightly older than M\,67.
In such CPDs stars of a given mass move upward on almost vertical lines as they 
spin down steadily with age \citep{Barnes16}.
The solar-mass stars, defined as those with $0.60 \leq (\bv)_0 \leq 0.70$
(and ignoring the outlier\footnote{This star has a comparably bright neighbour, EPIC 211411722, at a distance of $14.8''$ (see Table\,1).} 
IDW = 12030 = EPIC 211411477 at $P$ = 31.2\,d) 
have a mean rotation period of 25.8\,d 
(median = 26\,d; standard deviation $\sigma$ = 1.3\,d).
At the current level of precision, this value is indistinguishable from the 
solar rotation period of 26.1\,d (Donahue et al 1996).

It should be noted that the bluer stars of our sample have systematically 
smaller variability amplitudes (in agreement with prior open cluster results), 
indicating that they have smaller fractional spotted areas, while the reddest 
stars are variable enough to be detectable even from the ground. 
Stars with spectral types F8-G0 are not detectably spotted. 
Similar behaviors were noticed in the younger clusters NGC\,6819 and NGC\,6811
\citep[see][]{Meibom11, Meibom15}.

We have reported rotation periods for 20 stars of the 106 GLM15 member 
stars outside the area of the central K2 superstamp.
These are undoubtedly the M\,67 stars with both relatively large spot groups, 
and also those with rotational axes favorably inclined w.r.t. the earth. 
The remainder of the 106 members 
include stars where spot evolution does not allow good period determination, 
those that are either just unfavorably inclined, and
those that had insufficiently asymmetric spot distributions and/or 
smaller-than-detectable spot sizes during the K2 observations.

There are half a dozen additional periodic stars with lower variability levels
among the 106 members. 
Although their light curves are less convincing than the ones reported here
(at the current levels of light curve correction), they lie in the 
{\em same regions} of the CMD and CPD currently occupied by our sample.
This fact informs us that the CPD displayed in Fig.\,3 is robust against the 
particular choice of variables displayed.
Other researchers might construct M\,67 rotator samples using slightly 
different choices for individual stars, and may choose different data reduction 
strategies.
However, we opine that the CPD presented here is unlikely to be 
altered significantly by such choices.

Finally, we note that the $P = P(t, M)$ surface for M\,67 is expected to be 
intrinsically thinner than, say, that of the 2.5\,Gyr cluster NGC\,6819
because rotational evolution is highly convergent 
\citep[e.g.][]{Kawaler88, Barnes10a} under normal circumstances i.e. for 
stars that are not located in close binary systems or are otherwise 
pathological. 
The measured surface presented here is wider. 
We attribute this scatter to the long rotation periods of the sample stars 
relative to the K2 observing baseline. 
In contrast, the availability of multiple quarters of {\em Kepler} data for
NGC\,6819 enabled multiple determinations of each star's period 
\citep[see][]{Meibom15}, 
and the calculation of a corresponding mean stellar rotation period.

\subsection{Rotational age for the cluster}

\begin{figure}
\includegraphics[angle=-90,trim=0cm 3cm 0cm 0cm,clip=true,width=100mm]{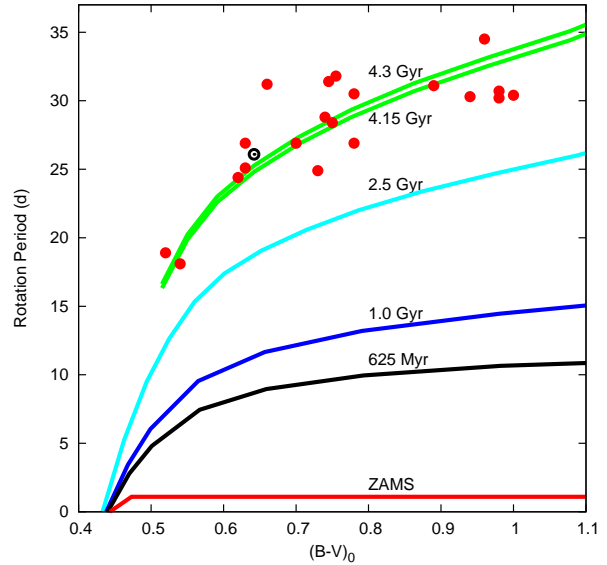}
\caption{Color-period diagram for M\,67, together with rotational isochrones 
for the mean (4.15\,Gyr) and median (4.3\,Gyr) ages, constructed using the 
model of \citet{Barnes10}. 
The current Sun is located approximately $1 \sigma$ above these isochrones. 
The majority of the data scatter about the isochrone is observational. 
We also display isochrones for certain younger ages for which the model has 
been tested against key cluster observations.
}
\end{figure}

Clearly all these stars, with- or without determined rotation periods, have a 
single age equal to the cluster age, $\sim$4\,Gyr 
\citep{Demarque92, VandenBerg04, Bellini10}. 
However, by treating the individual measured stars independently, i.e. as
field stars, each sampling the cluster age, we can examine the extent to which
gyrochronology yields the same age for individual cluster stars, and the
uncertainties with which ages for similar field stars might be derivable.

We therefore proceed to derive rotational ages, $t_i$, for the individual 
periodic rotators in M\,67 from the periods, $P_i$. 
Various models may be used for this purpose, beginning with those of 
Endal \& Sofia (1981) for solar-mass stars.
We use the relationship
\begin{equation}
      t = \frac{\tau}{k_C}{\rm ln} \left(\frac{P}{P_0}\right) 
        + \frac{k_I}{2 \tau} \left(P^2 - P_0^2\right)
\end{equation}
from B10, because of its prior success relative to other models in describing 
similar observations in a series of younger open clusters, including most 
recently, the 400\,Myr-old cluster M\,48 \citep{Barnes15}, and the 
2.5\,Gyr-old cluster NGC\,6819 \citep{Meibom15}\footnote{\citet{Lanzafame15} \citep[c.f.]{Brown14} have also shown that this model is better than others at describing the mass dependence of rotation. By including the turnover timescale, and therefore the Rossby number, {\em Ro}, this model also connects to chromospheric activity work by \citet{Noyes84}, and a body of dynamo-related work going back to at least \citet{Durney78}.}.
Here $P_0 = 1.1$\,d, and $k_C = 0.646$\,{\rm d/Myr}, $k_I = 452$\,{\rm Myr/d} 
are two dimensionless constants, retained unchanged from B10. 
(The dispersion from the range (0.12-3.4\,d) of possible initial rotation 
 periods is negligible by the age of M\,67.)
The convective turnover timescales, $\tau$, for each star were obtained by
interpolating the (global) values in column\,5 from Table\,1 in 
\citet{Barnes10a}\footnote{Note that it is possible to use other turnover timescales because all scale well with one another, thus leaving the power-law dependencies unchanged. However, the use of different timescales {\em requires recalibration of the constants} $k_C$ and $k_I$ to ensure that the rotational evolution model still describes all the open cluster data accurately.} 
from the corresponding $(\bv)_0$ colors in GLM15.
We have made one update; the rotational isochrone now takes into account the 
(small) blueward evolution of stars between the ZAMS and solar age. 

\begin{figure}
\includegraphics[angle=-90,width=\columnwidth]{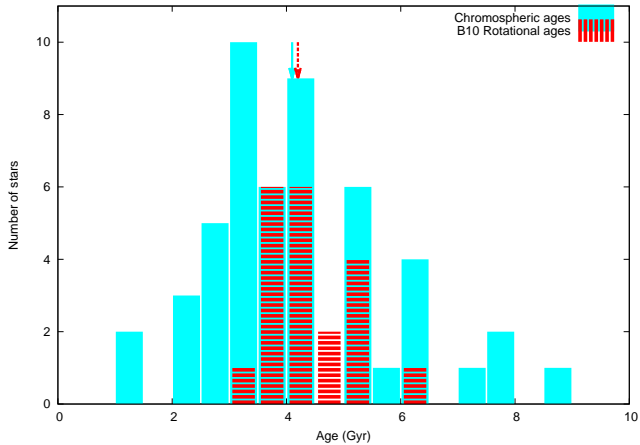}
\caption{\label{figure4}Histogram of gyrochronology ages (red) for the individual M\,67 stars.
All stars but one have $ 3.4\,{\rm Gyr}< {\rm Age}< 5.3\,{\rm Gyr}$, with
a median age of 4.28\,Gyr (mean = 4.15\,Gyr), and standard deviation of 
$0.7$\,Gyr (= 17\%). 
The equivalent MH08 chromospheric age distribution for the {\em single cluster 
members} of \citet{Giampapa06} is also displayed for comparison. 
These ages have a median of 3.95\,Gyr (mean = 4.2\,Gyr), and a standard 
deviation of 1.6\,Gyr (= 38\%). 
The arrows indicate the final chromospheric (4.1\,Gyr) and gyrochronology 
(4.2\,Gyr) ages for M\,67 (see text).
}
\end{figure}

The relevant rotational isochrones (sometimes called gyrochrones) are 
displayed in Fig.\,\ref{figure4}.
The mass dependence of the observations at the cluster age is clearly
described satisfactorily by the mean and median rotational isochrones, 
indicating a rotational age of 4.2\,Gyr (more below).
We also display the initial condition used for the ZAMS and rotational 
isochrones for younger open clusters of representative ages where the 
B10 model provides good fits to the data.
Because the rotation period evolution of stars in this diagram is almost 
exactly vertical 
(i.e. the color of a main sequence star hardly changes with age), 
this CPD succinctly displays how the rotation period for a
field star can be used to derive its age.

The histogram of the rotational ages for the M\,67 stars is displayed with red 
stripes in Fig.\,5.
We obtain a distribution peaked at $\sim$4\,Gyr, with all M\,67 periodic stars 
but one located in the interval $ 3.4\,{\rm Gyr}< {\rm Age}< 5.3\,{\rm Gyr}$, 
bracketing the mean- and median cluster ages. 
A (B10) rotational age for M\,67 outside this range is essentially excluded.
The single outlier is the star with $P = 31.2$\,d that is only slightly redder 
than the Sun (see also Table\,1).
The mean and median values are respectively, 4.15\,Gyr and 4.28\,Gyr, 
leading us to quote 4.2\,Gyr as the B10 rotational age of the cluster. 
The standard deviation of the individual age measurements is 0.7\,Gyr (= 17\%), 
which we take to represent the uncertainty that one might obtain for an FGK 
field star of similar age and metallicity from observations of similar quality\footnote{This is likely an underestimate. For instance, without a group of cluster stars to provide subtle guidance, the period derived for a field star with two symmetric spot groups would be half the true period, and the corresponding age would be 70\% of the true age.}.
The comparison between the B10 model and the measured data points gives 
$\chi_{\rm red}^2 = 2.35$. This implies that we are definitely not overfitting 
the M\,67 CPD. On the other hand, if the model is a good one, this value could 
be interpreted to mean that the period errors are slightly underestimated.

In this context, it should also be mentioned that there is a long history of
rotational stellar models beginning with \citet{Endal76}, 
whose details and predictions are beyond the scope of this paper. 
Post-2010 alternatives to the model presented here include
\citet{Lanzafame15, Gallet15, Johnstone15, 
Matt15, Brown14, Epstein14,Spada11}.
The models will undoubtedly be tested against these data in due course.

Let us now briefly consider the cluster as a whole.
The standard error on the mean cluster age is 
$0.7/\sqrt{19} = 0.16$\,Gyr.
But systematic errors could also add to the uncertainty in the rotational age 
of the cluster.
For instance, a non-solar metallicity could conceivably affect the rotational 
age of the cluster, especially as gyrochronology is currently calibrated only 
for solar-metallicity stars. 
Fortunately, this is likely negligible because high-resolution spectroscopic
studies of M\,67 \citep[(e.g.][]{Jacobson11,Randich06} find
[Fe/H] values essentially indinguishable from solar.
Finally, there is the reddening uncertainty.
Again, this cannot be very large because the cluster itself is off the 
Galactic plane, and thus barely reddened, with $E(\bv) = 0.04$.
Cautious investigators such as \citet{Brucalassi04} consider values 
as distant as $E(\bv)=0.02$. 
However, \citet{Taylor07} insists on a reddening uncertainty of only 0.004\,mag,
and \citet{VandenBerg04} believe the reddening to be established
to an uncertainty of only $\pm$0.005. 
Such a change would perturb the cluster gyrochronology age by only 
$\pm$0.05\,Gyr.
Adding this in quadrature to the standard error on the mean gives a total
uncertainty of $0.17$\,Gyr, which we round to 0.2\,Gyr and adopt as the
uncertainty on the B10 rotational age of M\,67, listing it finally as 
4.2$\pm$0.2\,Gyr.

The results reported here accord well with prior work on M\,67.
\citet{Giampapa06} have studied the chromospheric emission of M\,67 stars,
confirming a mean emission similar to that of the Sun, but with greater
excursions from the mean. We have (re-)calculated stellar ages from their 
measurements using the activity-age calibration of 
\citet{Mamajek08}, retaining only the 50 single cluster members of 
GLM15, and present this histogram for comparison with the rotational ages in 
Fig.\,5.
This chromospheric age distribution has mean and median values of 4.2\,Gyr 
and 3.95\,Gyr respectively, and a standard deviation of 1.6\,Gyr (= 38\%), 
in agreement with prior knowledge.
The chromospheric age of the cluster is therefore $4.1 \pm 0.23$\,Gyr, where 
the uncertainty quoted is solely the standard error on the mean.
This value is identical to the rotational age to within the uncertainties.

\section{Conclusions}

The rotation periods of cool (FGK) single member stars in M\,67 define a 
sequence in the color-period diagram reminiscent of that discovered first
in the Hyades open cluster. 
The Sun lies marginally above the sequence in the M\,67 CPD, and on the same 
$P = P (t, M)$ surface defined by prior open cluster observations. 
This fact strengthens the solar-stellar connection.

The location of the rotational sequence in the color-period diagram is in 
agreement with the predictions of prior (semi-)empirical rotational models, 
implying that reliable rotational ages can be derived for solar-metallicity 
dwarfs upto solar age, and likely upto the main sequence turnoff.

The gyro-ages of the individual cluster members have a standard deviation
of {0.7}\,Gyr (= 17\%), suggesting that similar age errors are attainable
with K2 data for equivalent field stars (e.g. planet hosts), 
provided their surfaces are sufficiently asymmetrically spotted, and spot 
evolution is not severe-enough to prevent period determination.
The gyro-age for M\,67 as a whole is 4.2$\pm$0.2\,Gyr, 
with the uncertainty originating primarily in the period determination errors, 
and to a lesser extent from reddening and metallicity uncertainties.

Finally, we note that the variability of sun-like stars is at a similar level 
as that of the (quiet-to-active) Sun, and that two major spot groups at widely
different longitudes are evident in 70\% of the light curves in the sample
reported here.

\acknowledgments

We acknowledge the anonymous referee for a careful and helpful referee report.
We are grateful to the German and US tax payers for supporting this work,
and to NASA and K2 for making the data publicly available.

{\it Facilities:} \facility{Kepler}, \facility{K2}.

\appendix

\begin{figure}
\includegraphics[angle=0,width=\columnwidth]{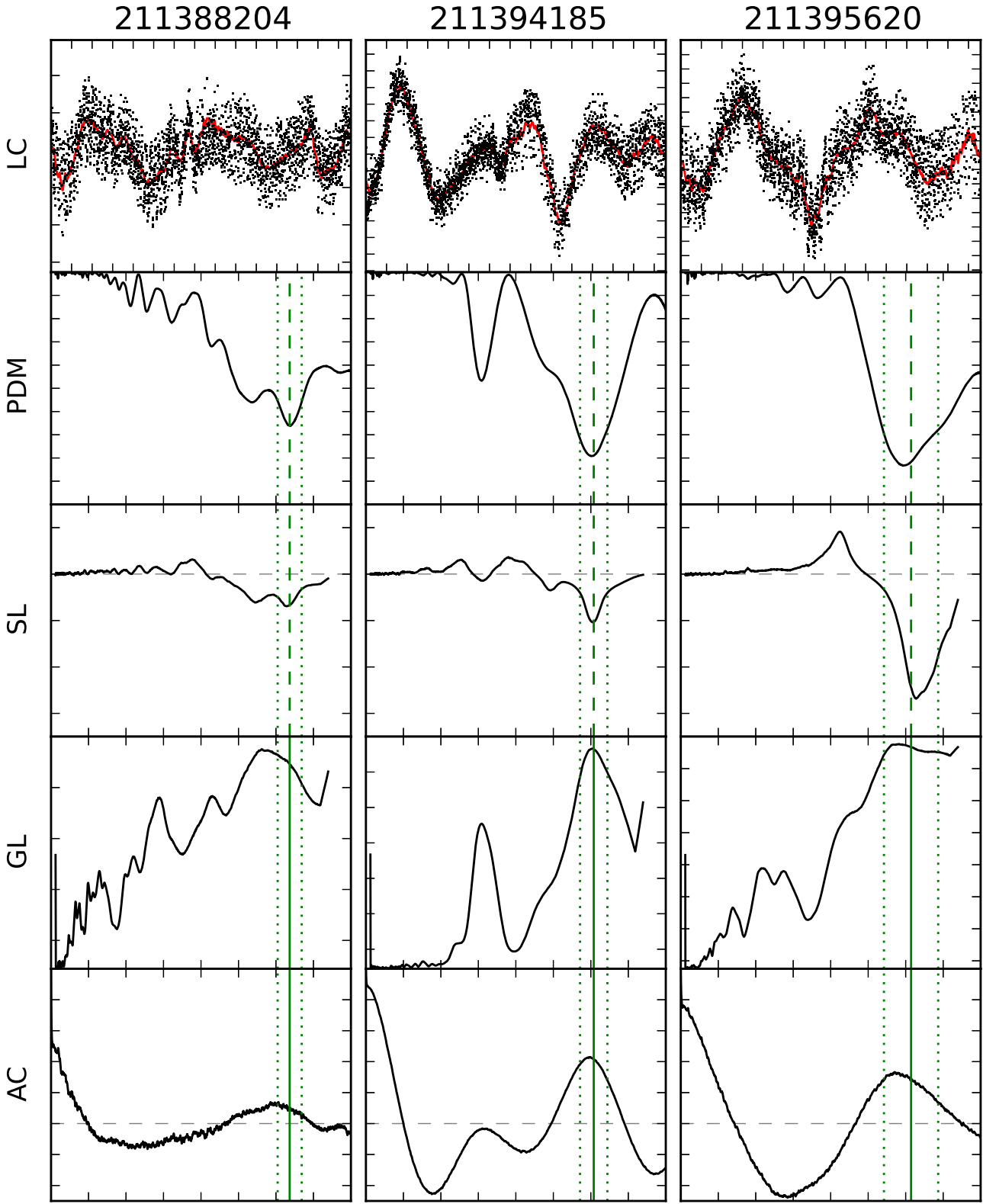}
\caption{
Light curves and results from four period analysis methods, with the abscissae marked at 5\,d intervals, and the final period indicated. The ordinate for the light curves indicates 1\,mmag intervals, while those for the PDM, SL, GL, and AC panels give the $\Theta$ statistic ($0\,{\rm to}\, 1$), the deviation of the measured string length ($-35\,{\rm to}\,15$), the natural logarithm of the posterior probability of a given period (with upper limit $P=1$), and the autocorrelation of the time series ($-0.5\,{\rm to}\,1$) respectively.  
}
\end{figure}

\begin{figure}
\includegraphics[angle=0,width=\columnwidth]{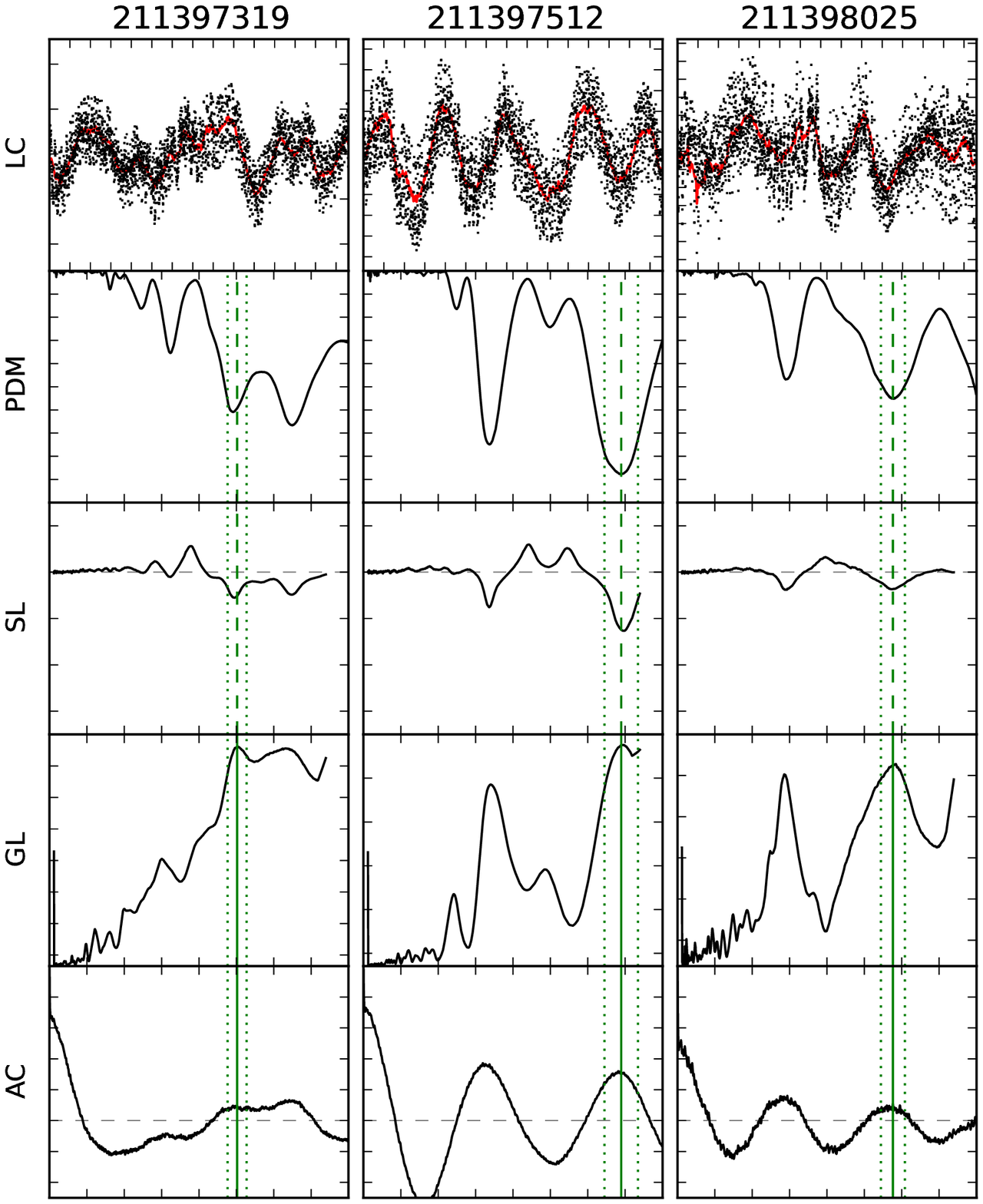}
\caption{
Light curves and results from four period analysis methods, with the abscissae marked at 5\,d intervals, and the final period indicated. The ordinate for the light curves indicates 1\,mmag intervals, while those for the PDM, SL, GL, and AC panels give the $\Theta$ statistic ($0\,{\rm to}\, 1$), the deviation of the measured string length ($-35\,{\rm to}\,15$), the natural logarithm of the posterior probability of a given period (with upper limit $P=1$), and the autocorrelation of the time series ($-0.5\,{\rm to}\,1$) respectively.  
}
\end{figure}

\begin{figure}
\includegraphics[angle=0,width=\columnwidth]{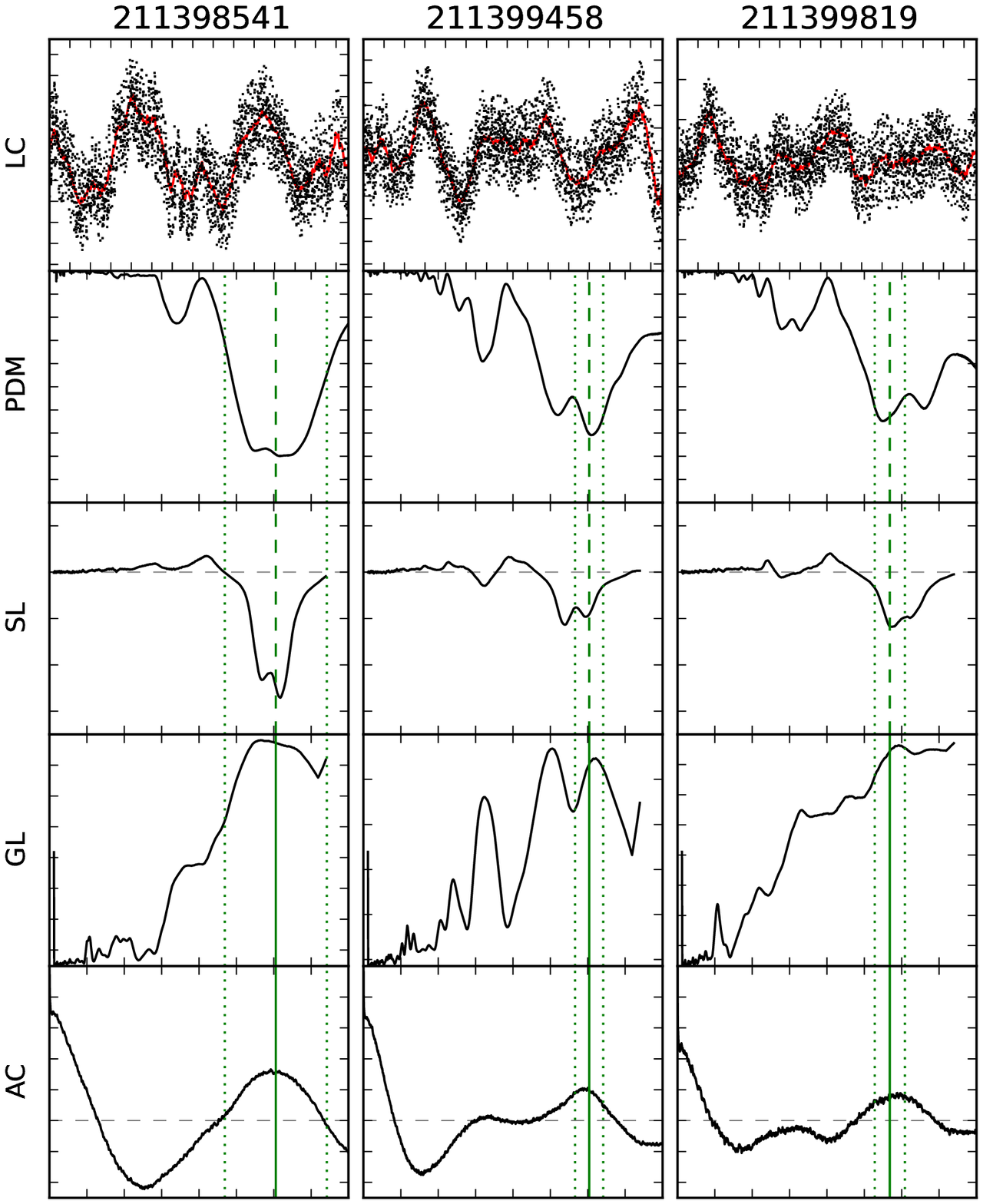}
\caption{
Light curves and results from four period analysis methods, with the abscissae marked at 5\,d intervals, and the final period indicated. The ordinate for the light curves indicates 1\,mmag intervals, while those for the PDM, SL, GL, and AC panels give the $\Theta$ statistic ($0\,{\rm to}\, 1$), the deviation of the measured string length ($-35\,{\rm to}\,15$), the natural logarithm of the posterior probability of a given period (with upper limit $P=1$), and the autocorrelation of the time series ($-0.5\,{\rm to}\,1$) respectively.  
}
\end{figure}

\begin{figure}
\includegraphics[angle=0,width=\columnwidth]{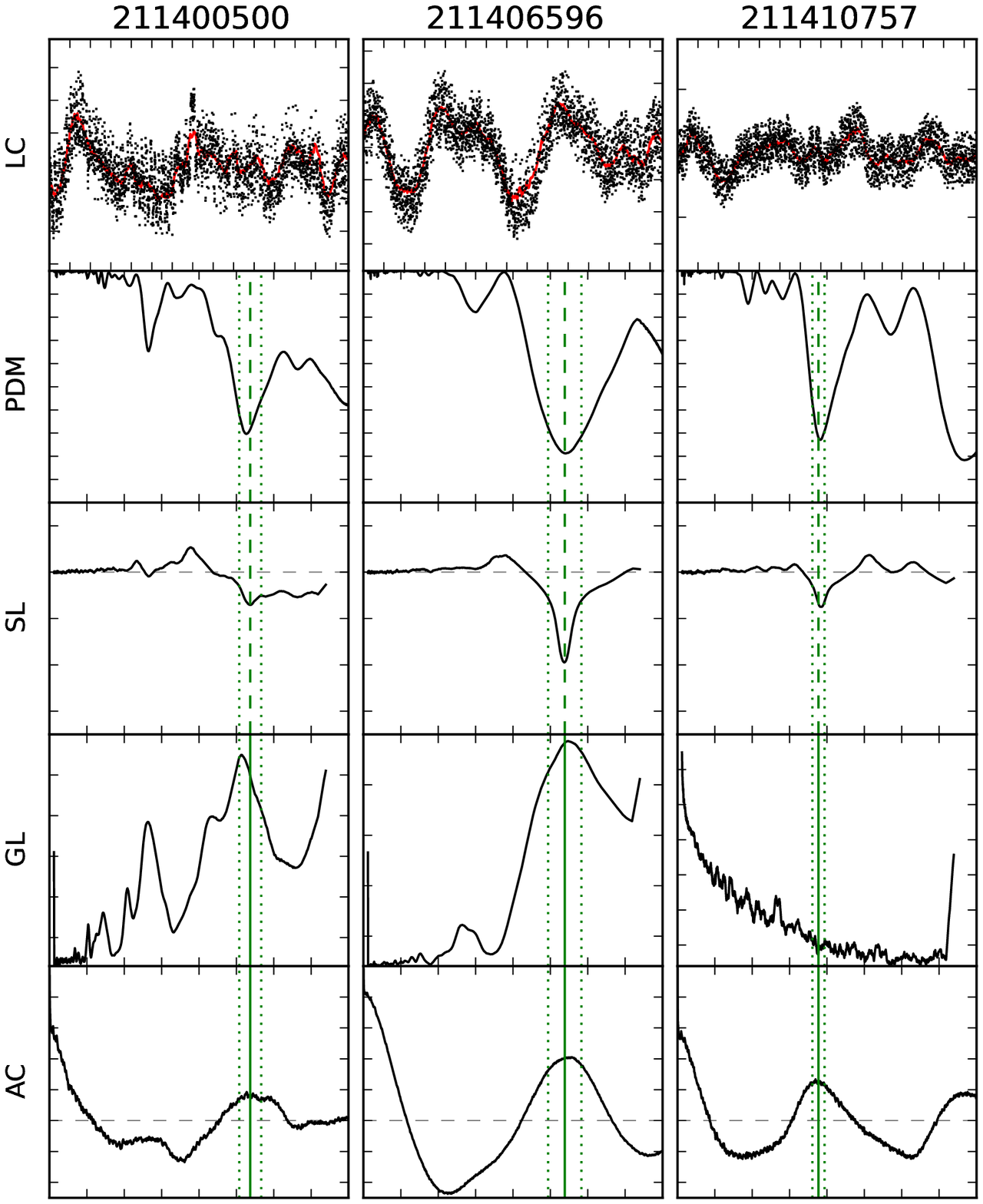}
\caption{
Light curves and results from four period analysis methods, with the abscissae marked at 5\,d intervals, and the final period indicated. The ordinate for the light curves indicates 1\,mmag intervals, while those for the PDM, SL, GL, and AC panels give the $\Theta$ statistic ($0\,{\rm to}\, 1$), the deviation of the measured string length ($-35\,{\rm to}\,15$), the natural logarithm of the posterior probability of a given period (with upper limit $P=1$), and the autocorrelation of the time series ($-0.5\,{\rm to}\,1$) respectively.  
}
\end{figure}

\begin{figure}
\includegraphics[angle=0,width=\columnwidth]{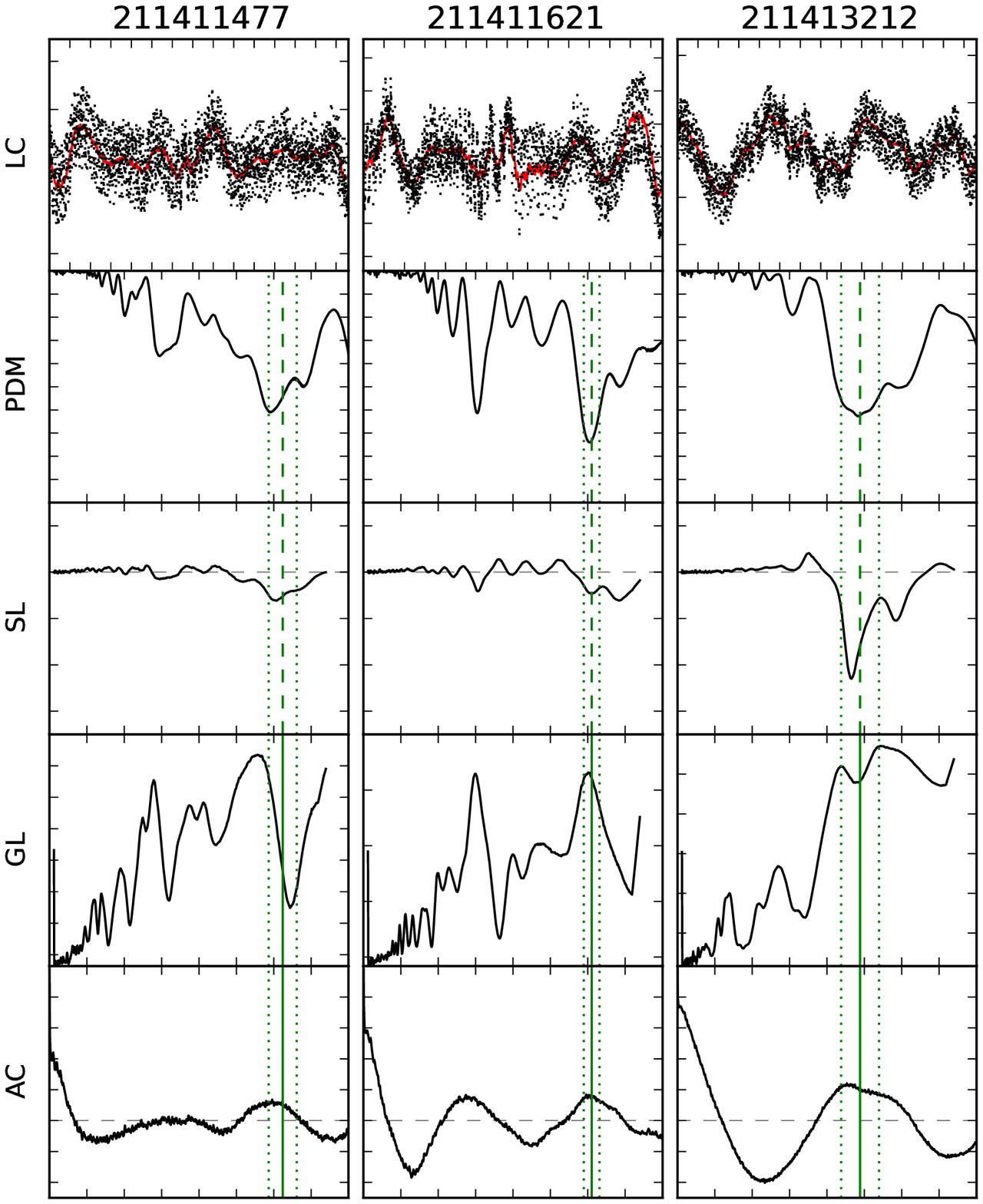}
\caption{
Light curves and results from four period analysis methods, with the abscissae marked at 5\,d intervals, and the final period indicated. The ordinate for the light curves indicates 1\,mmag intervals, while those for the PDM, SL, GL, and AC panels give the $\Theta$ statistic ($0\,{\rm to}\, 1$), the deviation of the measured string length ($-35\,{\rm to}\,15$), the natural logarithm of the posterior probability of a given period (with upper limit $P=1$), and the autocorrelation of the time series ($-0.5\,{\rm to}\,1$) respectively.  
}
\end{figure}

\begin{figure}
\includegraphics[angle=0,width=\columnwidth]{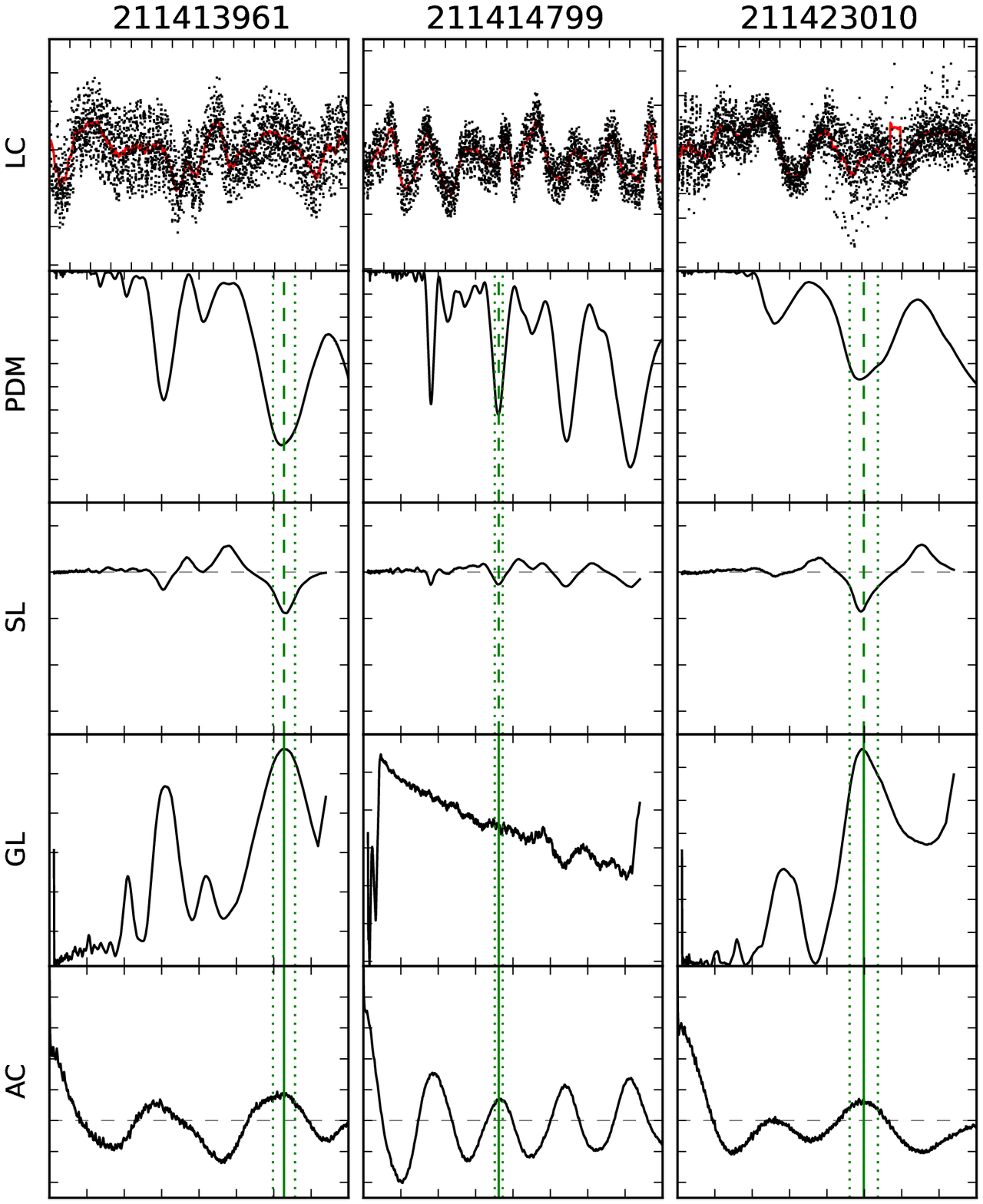}
\caption{
Light curves and results from four period analysis methods, with the abscissae marked at 5\,d intervals, and the final period indicated. The ordinate for the light curves indicates 1\,mmag intervals, while those for the PDM, SL, GL, and AC panels give the $\Theta$ statistic ($0\,{\rm to}\, 1$), the deviation of the measured string length ($-35\,{\rm to}\,15$), the natural logarithm of the posterior probability of a given period (with upper limit $P=1$), and the autocorrelation of the time series ($-0.5\,{\rm to}\,1$) respectively.  
}
\end{figure}

\begin{figure}
\includegraphics[angle=0,width=12cm]{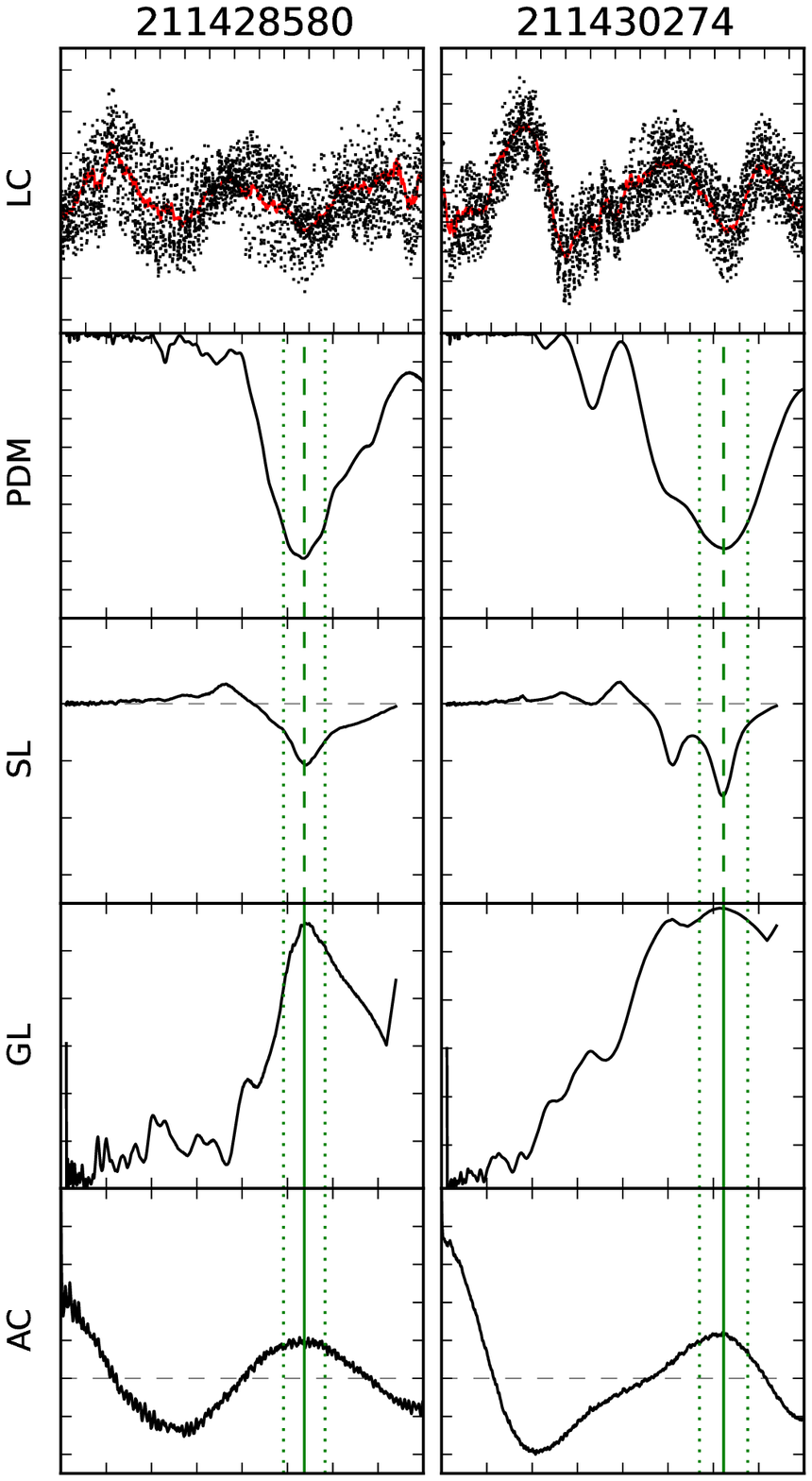}
\caption{
Light curves and results from four period analysis methods, with the abscissae marked at 5\,d intervals, and the final period indicated. The ordinate for the light curves indicates 1\,mmag intervals, while those for the PDM, SL, GL, and AC panels give the $\Theta$ statistic ($0\,{\rm to}\, 1$), the deviation of the measured string length ($-35\,{\rm to}\,15$), the natural logarithm of the posterior probability of a given period (with upper limit $P=1$), and the autocorrelation of the time series ($-0.5\,{\rm to}\,1$) respectively.  
}
\end{figure}


\begin{figure}
\includegraphics[angle=0,width=\columnwidth]{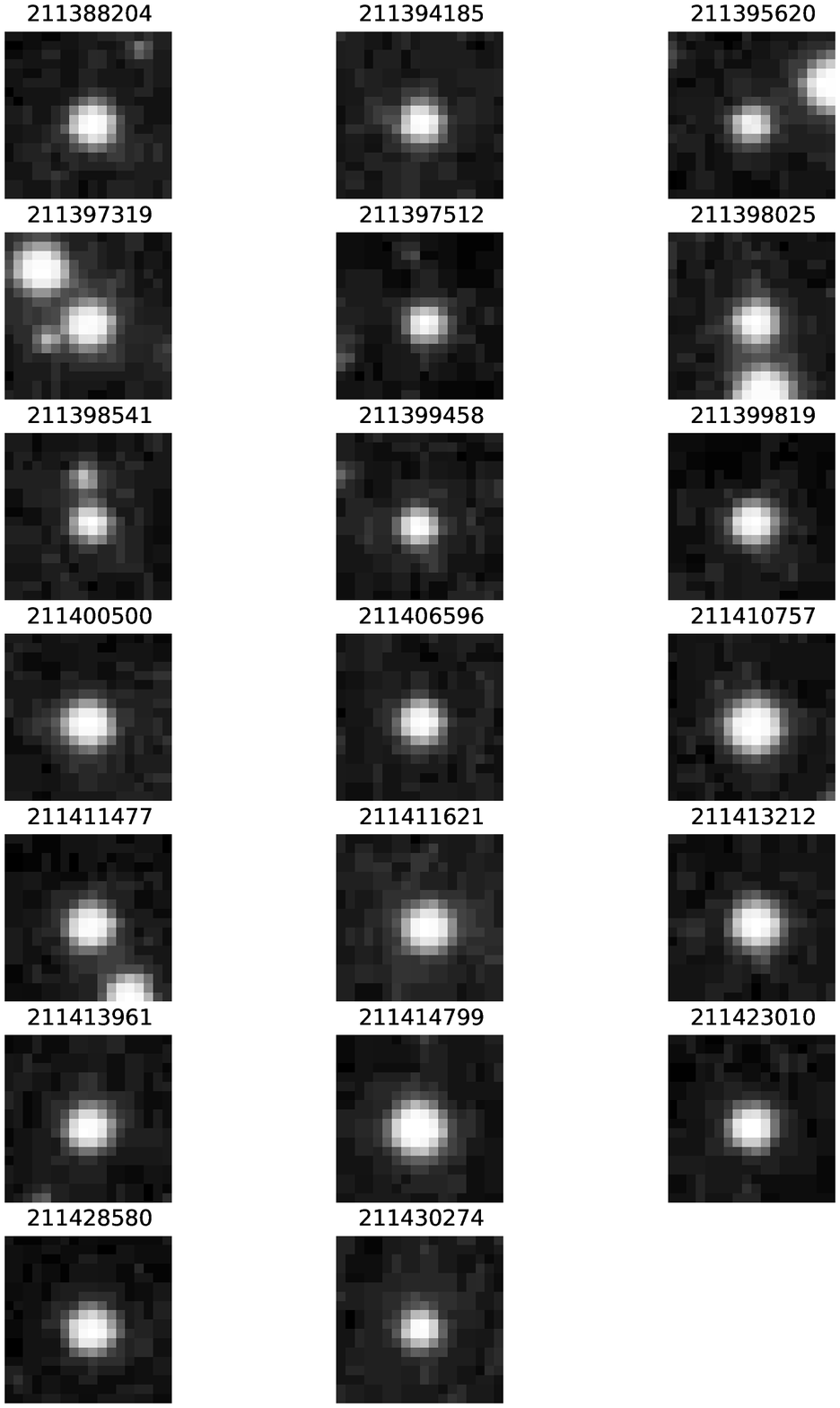}
\caption{$30 \arcsec \times 30 \arcsec$ cutouts from the digitized sky survey (DSS1) image of the M\,67 cluster region.
}
\end{figure}

\begin{figure}
\includegraphics[angle=0,width=\columnwidth]{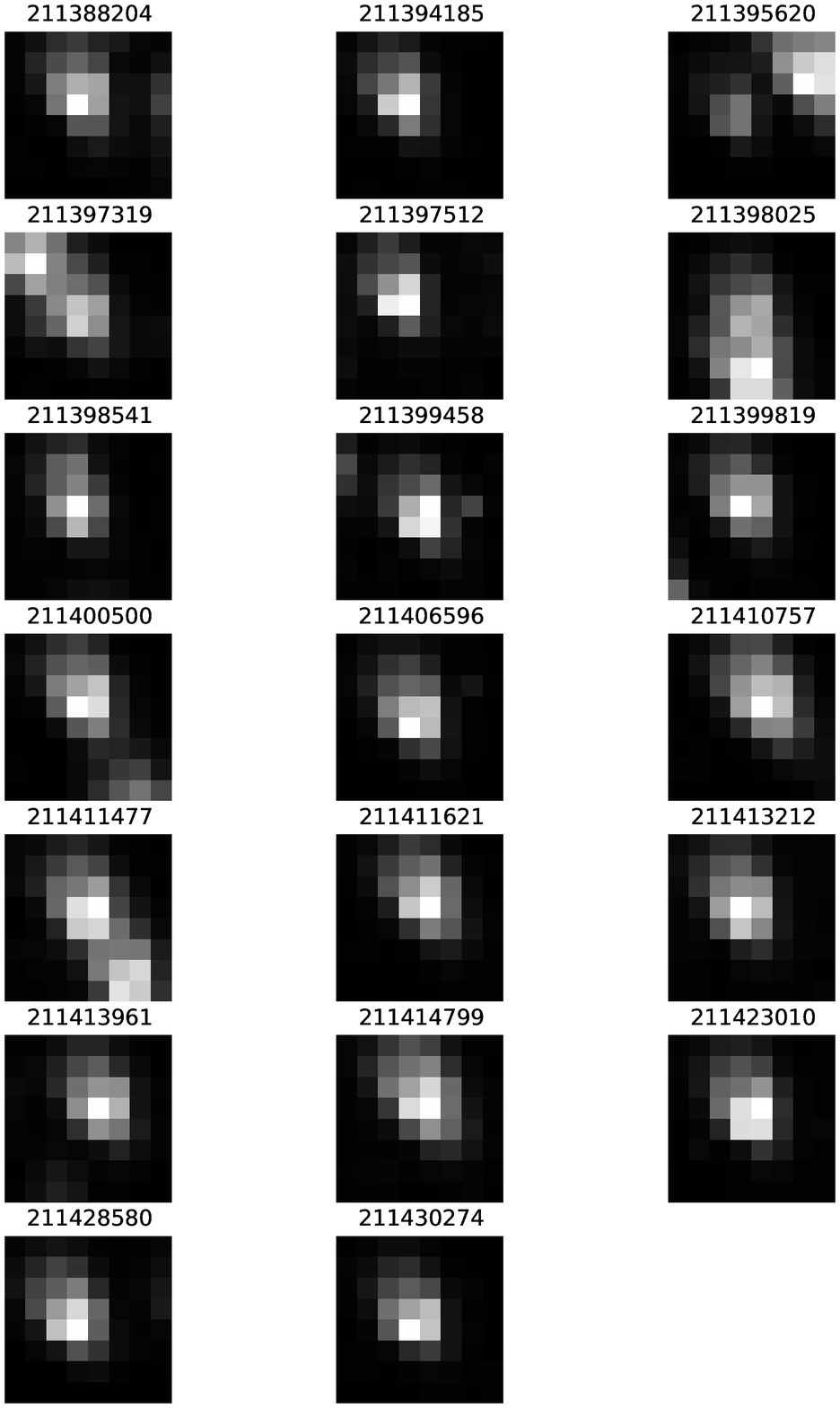}
\caption{$30 \arcsec \times 30 \arcsec$ cutouts from the final full-frame K2 image of the M\,67 cluster region.
}
\end{figure}






\bibliographystyle{apj}
\bibliography{m67}

\end{document}